\documentclass[useAMS,usenatbib,usegraphicx]{mn2e}

\usepackage{aas_macros} 

\newcommand{\csixty}{C\ensuremath{_{60}}}
\newcommand{\arii}{[Ar\,\textsc{ii}]}
\newcommand{\neii}{[Ne\,\textsc{ii}]}
\newcommand{\neiii}{[Ne\,\textsc{iii}]}
\newcommand{\Siii}{[Si\,\textsc{ii}]}
\newcommand{\Suiii}{[S\,\textsc{iii}]}

\title[C$_{60}$ in young stellar and other objects]{Detection of C$_{60}$ in embedded young stellar objects, a Herbig Ae/Be star and an unusual post-AGB star.}

\author[K.~R.~G. Roberts, K.~T. Smith and P.~J. Sarre]{Kyle R.~G. Roberts\thanks{\texttt{pcxkr@nottingham.ac.uk}}, Keith T. Smith\thanks{\texttt{keith.smith@nottingham.ac.uk; present address - Royal Astronomical Society, Burlington House, Piccadilly, London W1J 0BQ, U.K.
 }} and Peter J. Sarre\thanks{\texttt{peter.sarre@nottingham.ac.uk}}\\
School of Chemistry, The University of Nottingham, University Park, Nottingham NG7 2RD, U.K.}

\begin{document}

\date{Accepted 2012 January 12. Received 2012 January 12; in original form 2011 September 15}

\pagerange{\pageref{firstpage}--\pageref{lastpage}} \pubyear{2011}

\maketitle

\label{firstpage}

\begin{abstract}
The first detection of the \csixty\ (Buckminsterfullerene) molecule in massive embedded young stellar objects (YSOs) is reported. Observations with \emph{Spitzer} IRS reveal the presence of \csixty\ in YSOs ISOGAL-P J174639.6-284126 and SSTGC~372630 in the Central Molecular Zone in the Galactic centre, and in a YSO candidate, 2MASS~J06314796+0419381, in the Rosette nebula. The first detection of \csixty\ in a Herbig Ae/Be star, HD~97300, is also reported.  These observations extend the range of astrophysical environments in which \csixty\ is found to YSOs and a pre-main sequence star. \csixty\ excitation and formation mechanisms are discussed in the context of these results, together with its presence and processes in post-AGB objects such as HR~4049.
\end{abstract}


\section{Introduction}
\label{sec:intro}

Although considered a candidate carrier for a few infrared astronomical emission features, as noted in \citet{Werner2004a} and discussed by \citet{Sellgren2007}, confirmation of C$_{60}$ together with C$_{70}$ in the planetary nebula Tc~1 \citep{Cami2010} and C$_{60}$ in the reflection nebulae NGC~7023 and NGC~2023 \citep{Sellgren2010,Sellgren2011} has only recently been achieved. Further discoveries of neutral \csixty\ have since been reported in planetary nebulae (PNe) in the Milky Way and the Magellanic Clouds \citep{Garcia2010,Garcia2011a}, weakly H-deficient R~Coronae Borealis-type stars \citep{Garcia2011}, a protoplanetary nebula \citep{Zhang2011}, across the `veil' region of the Orion Nebula \citep{Rubin2011}, in several post-AGB objects \citep{Gielen2011,Gielen2011a} and possibly in the binary XX~Oph \citep{Evans2011}. There is therefore now unambigious evidence for neutral \csixty\ in objects in the later stages of stellar evolution and in the interstellar medium.

 We report here the first detection of \csixty\ in two young stellar objects (YSOs) located in the Central Molecular Zone (CMZ) of the Milky Way, in a YSO candidate in the Rosette nebula and in a Herbig Ae/Be star. Taken with earlier reports, these findings show that neutral \csixty\ exists over a very wide spectrum of stellar and interstellar evolution, encompassing star forming regions and young stars, mass-losing evolved stars, protoplanetary and planetary nebulae, and the interstellar medium. We also report detection of \csixty\ in the post-AGB star HR~4049 which, like HD~52961 in which \csixty\ is known \citep{Gielen2011a}, is unusual in displaying the rare infrared emission signatures of nanodiamonds.

This paper is organised as follows: background to current issues associated with astrophysical \csixty\ is presented in section~\ref{sec:background}, details of the \emph{Spitzer} observations and data reduction in section~\ref{sec:obs} and the classification of the YSOs\footnote{We choose to use the term YSO to describe the embedded YSOs rather than include the more evolved Herbig~Ae/Be star.} and the YSO candidate in section~\ref{sec:classification}.  Measurements and analysis of \csixty\ features are presented in section~\ref{sec:C60bands}. Excitation mechanisms and \csixty\ formation are discussed in section~\ref{sec:discuss} and the results are summarised in section~\ref{sec:summary}.

\section{Background}
\label{sec:background}


Astronomical conditions analogous to those of the early laboratory experiments in which \csixty\ is formed \citep{kroto1985,kratschmer1990} might have been expected to favour \csixty\ formation, with hydrogen-poor R~Coronae Borealis (RCB) stars being good candidates. However, this has not been borne out by observation \citep{Clayton1995} except in the case of one or possibly two less H-poor RCB stars \citep{Garcia2011}.  Alternatively, laser-induced decomposition of hydrogenated amorphous carbon (HAC) is known to produce fullerenes \citep{Scott1997} and HAC decomposition has been referred to as a possible astrophysical \csixty\ formation route \citep{moutou1999,Ehrenfreund2000,Sellgren2010,Garcia2010}.  Following identification of \csixty\ and C$_{70}$ in the particularly clean spectrum of Tc~1, \citet{Cami2010} suggested that these molecules formed efficiently in this object because the circumstellar environment was H-poor. However, in their paper on \csixty\ emission in PNe  \citet{Garcia2010} pointed out that neither the PN nor its compact core and the central star of Tc~1 are H-poor, and presented evidence in support of the idea that fullerenes are formed in astrophysical environents by decomposition of HAC \citep{Garcia2011a}. They suggested that \csixty\ and polycyclic aromatic hydrocarbons (PAHs) were likely to be formed together from the decomposition of HAC due to UV processing and energetic phenomena such as shocks, and interpreted the deficiency of PAHs in Tc~1 in terms of longer survival times of fullerenes due to their high stability. \csixty\ formation issues are further explored in papers by \cite{Cami2011} and \cite{Garcia2011b}, with the challenges being extended further by the detection of \csixty\ in mixed-chemistry post-AGB stars \citep{Gielen2011} and reported for HR~4049 in this paper.

The phase (gas or solid) and excitation mechanism of \csixty\ in astrophysical sources is not firmly established and may vary between environments. From their analysis of spectral data for Tc~1 and particularly the relatively low vibrational temperature of $\sim$\,330~K that was inferred, \citet{Cami2010} suggested that \csixty\ in Tc~1 is attached to, and in thermal equilibrium with, solid material such as carbonaceous grains. In contrast, \citet{Sellgren2010} have found the vibrational band intensity ratios in NGC~7023  to be consistent with high-energy ($\sim$\, 10~eV) UV photo-excitation of gas-phase \csixty. At present we are not aware of predicted band strengths based on other possible excitation mechanisms. The discovery of \csixty\ in YSOs and a Herbig Ae/Be star provides additional environments with which proposed formation and excitation mechanisms can be evaluated.

\section{Observations and data reduction}
\label{sec:obs}

A search was undertaken for \csixty\ emission bands in spectra recorded with the Infra-Red Spectrograph (IRS; \citealt{Houck2004}) of the \emph{Spitzer} Space Telescope \citep{Werner2004}. We interrogated the \emph{Spitzer} Heritage Archive (SHA)\footnote{\texttt{http://archive.spitzer.caltech.edu}} for objects observed at the wavelengths of the \csixty\ emission bands, concentrating on pre-main sequence objects and examining several hundred spectra. Targets were selected by visual inspection of pipeline-reduced data in the spectral region around the 18.9\,\micron\ \csixty\ band, which is relatively strong and free from blending with atomic lines and PAH emission bands. Data for each object with a potential 18.9\,\micron\ detection were retrieved from the archive, together with additional targets selected for comparison. Observational details and photometry for the objects found to contain \csixty\ are given in table \ref{tab:targets}.

All of the targets were observed using the IRS Short-High (SH) module which covers the wavelength region 9.9-19.6\,\micron\ with resolving power $R\equiv\frac{\lambda}{\delta\lambda}\sim600$. Some targets were also observed with the Short-Low (SL) module (5.2-14.5\,\micron, $R\sim$\,60-120), and/or the Long-High (LH) module (18.7-37.2\,\micron, $R\sim600$).

\begin{table*}
\caption{Coordinates and photometry for the targets with \csixty\ emission bands. Right ascension, declination, and near-infrared magnitudes are taken from the 2MASS catalogue \citep{Skrutskie2006}; mid-infrared magnitudes (where available) are from the \emph{Spitzer} IRAC survey of the Galactic centre \citep{Ramirez2008}. \emph{Spitzer} programme numbers and Principal Investigator (PI) names are are given for the IRS observations used in this study.}
 \begin{tabular}{lccccccccl}
  \hline
  Name & RA & Dec & $J$ & $K$ & [3.6] & [4.5] & [5.8] & [8.0] & Programme (PI)\\
  \hline
  Embedded YSOs:\\
  ISOGAL-P J174639.6-284126 & 17:46:39.60 & $-$28:41:27.0 & $>13.8$ & 12.95 & 10.28 & 8.83 & 7.38 & 5.58 & 40230 (Ram\'irez)\\
  SSTGC 372630 & 17:44:42.76 & $-$29:23:16.2 & $>16.0$ & 12.87 & 10.31 & 8.82 & 7.67 & 6.48 & 40230 (Ram\'irez)\\ 
  2MASS J06314796+0419381 & 06:31:47.96 & +04:19:38.2 & 14.01 & 10.67 & - & - & - & - & 50146 (Keane)\\ 
  \\
  Other targets:\\
  HD 97300 (Herbig Ae/Be star) & 11:09:50.03 & $-$76:36:47.7 & 7.64 & 7.15 & - & - & - & - & 2 (Houck)\\
  HD 52961 (post-AGB object) & 10:18:07.52 & $-$28:29:30.7 & 16.06 & 15.42 & - & - & - & - & 3274 (Van Winckel)\\
  HR 4049 (post-AGB object) & 07:03:39.63 & +10:46:13.1 & 6.32 & 5.53 & - & - & - & - & 93 (Cruikshank)\\
  \hline
 \end{tabular}
\label{tab:targets}
\end{table*}

Flux- and wavelength-calibrated Post Basic Calibrated Data produced by version S18.18.0 of the Spitzer Science Centre pipeline were obtained from the archive, with further analysis undertaken using \textsc{iraf}\footnote{IRAF is distributed by the National Optical Astronomy Observatories, which are operated by the Association of Universities for Research in Astronomy, Inc., under cooperative agreement with the National Science Foundation.}. The observations were background-subtracted after inspection of 2MASS $K$-band images \citep{Skrutskie2006} for the presence of contaminating sources. Where no such sources were present, either the background subtraction performed by the pipeline was used (SL), or all of the background regions were combined and manual subtraction was employed (SH and LH). Where such sources were found to be present, only those background regions without significant contamination were combined and the background was subtracted manually. Multiple exposures were combined to increase the signal-to-noise ratio.

For targets observed with more than one IRS module, the spectra were combined by scaling the flux in overlapping wavelength regions. Where these were in disagreement, the flux from the SH observations was preferred and the other fluxes scaled accordingly, because the SH module has a smaller field of view and thus samples the flux from the target more reliably. In the final spectra, the wavelength ranges used from each module were: 5.2--10 (SL), 10--19.5 (SH) and 19.5--36\,\micron\ (LH).

\section{Object classification}
\label{sec:classification}

Given the importance of determining that the YSOs discussed in this paper are embedded young objects, we review here the evidence that our two CMZ objects, ISOGAL-P~J174639.6-284126 and SSTGC~372630, are indeed YSOs and not dust-enshrouded post-AGB sources. Based on the available data, we designate our third (Rosette nebula) object, 2MASS~J06314796+0419381, as a candidate (but very likely) YSO.  The attribution to YSOs rests principally on the steep smooth rise in the spectral energy distribution (SED) with increasing wavelength, the presence of silicate and CO$_2$ ice absorption bands, the \neii\ 12.8\,\micron\ line and PAHs in emission, and the absence of a broad 30\,\micron\ emission feature\footnote{The 30\,\micron\ feature has been ascribed to MgS \citep{Goebel1985,Hony2002} but may, perhaps more probably, be due to carbonaceous material \citep{Zhang2009,Zhang2011} such as HAC \citep{Grishko2001}.} which is well correlated with the presence of \csixty\ in post-AGB objects. The Herbig Ae/Be star HD~97300 and the post-AGB star HR~4049 are well known and not discussed in this section.

\subsection{ISOGAL-P~J174639.6-284126}

\begin{figure}
\includegraphics[angle=-90,width=\columnwidth]{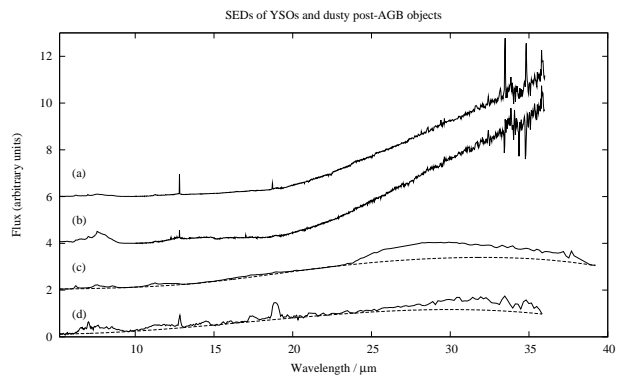}
	\caption[IRS spectra of ISOGAL-P J174639.6-284126 and SSTGC~372630 (in the CMZ) and comparison post-AGB objects.]{\emph{Spitzer} IRS spectra of the CMZ objects (a) ISOGAL-P J174639.6-284126 and (b) SSTGC~372630, illustrating the strong rise in the far-IR SED which is indicative of a YSO. The comparison traces are for (c) proto-planetary nebula IRAS~01005+7910 and (d) planetary nebula SMP SMC 016.  These show a slower rise in SED (indicated by the dashed lines) and a broad 30\,\micron\ emission feature commonly seen in \csixty-containing post-AGB objects.  The spectra are normalised to the flux of ISOGAL-P J174639.6-284126 at 22.5\,\micron\ and vertical offsets applied.}
	\label{fig:SEDrise}
\end{figure}

ISOGAL-P~J174639.6-284126 is located in the CMZ of the Milky Way (see table~\ref{tab:targets}). It was identified as a massive YSO by \citet{Felli2002} based on \emph{ISO} mid-infrared photometry.  Its \emph{Spitzer} IRS spectrum is shown in figure~\ref{fig:SEDrise} as trace (a) and illustrates a steep rise in SED with increasing wavelength, an SED peaking to the red of $\sim$35\,\micron\ being indicative of a young object \citep[e.g.][]{Simpson2011}. The figure shows that the SED of ISOGAL-P~J174639.6-284126 has a much faster rise at longer wavelengths than those of two sample post-AGB objects shown in traces (c) and (d). \citet{An2011} identified ISOGAL-P~J174639.6-284126 as a candidate YSO based on 2MASS and \emph{Spitzer} photometry, but excluded it from their final list of YSOs due to the lack of a $\sim15.4$\,\micron\ shoulder on the CO$_2$ ice absorption feature.  The presence of a shoulder at $\sim15.4$\,\micron, thought to be due to methanol-CO$_2$ ice complexes, was used as the YSO selection criterion principally to distinguish between YSOs and field stars behind molecular clouds. However, other authors \citep[e.g.][]{Seale2009,Simpson2011} have identified massive YSOs which do not have CO$_2$ ice absorption features, or have different peak wavelengths and profiles from those used by \citet{An2011} for selecting YSOs. Also, \cite{Ehrenfreund1999} have shown that photolysis of methanol-containing CO$_2$ ice mixtures decreases the strength of the shoulder on the CO$_2$ ice band; this could lead to disappearance of the shoulder from spectra of the more evolved YSOs, such as those associated with a compact or ultra-compact H\,\textsc{ii} region that are likely to have higher UV flux (see section~\ref{sec:evolution}).

\begin{figure}
\includegraphics[angle=-90,width=\columnwidth]{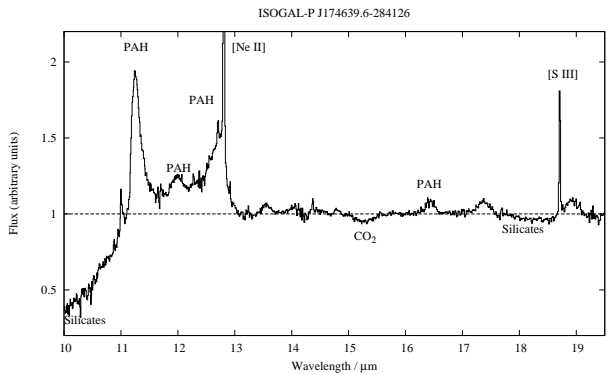}
	\caption[IRS spectrum from 5 to 19.5\,\micron\ of ISOGAL-P J174639.6-284126] {Continuum-normalised spectrum of ISOGAL-P J174639.6-284126 between 10 and 19.5\,\micron\ showing the presence of silicate and CO$_2$ ice absorption.}
	\label{fig:isogalcont}
\end{figure}

From figure~\ref{fig:SEDrise} it is clear that there is no 30\,\micron\ emission feature as seen in the \csixty-containing PPN IRAS~01005+7910 \citep{Zhang2011} and PN SMP SMC 016 \citep{Garcia2010}.\footnote{Spectra of PNe \citep{Bernard-Salas2009} and PPNe \citep{Volk2011} provide further examples.} Figure~\ref{fig:isogalcont} shows the continuum-normalised spectrum of ISOGAL-P~J174639.6-284126 between 10 and 19.5\,\micron\ with two silicate absorption bands and absorption by CO$_2$ ice indicated. Based on these spectral characteristics, together with the presence of PAH and \neii\ emission in this (and our other) YSOs which is generally absent in low-mass YSOs \citep{vandishoeck2004,Geers2009}, it is concluded that ISOGAL-P~J174639.6-284126 is a high-mass YSO.

\subsection{SSTGC~372630}
\label{sec:sstspectrum}

SSTGC~372630 is also located in the CMZ (see table~\ref{tab:targets}). Based on its SED and position in an infrared colour-magnitude diagram, it was identified as a massive YSO by \citet{Yusef-Zadeh2009}. Its \emph{Spitzer} IRS spectrum between 5 and 36\,\micron\ is given in figure~\ref{fig:SEDrise} - trace (b) and, like ISOGAL-P~J174639.6-284126, shows a steeply rising continuum in the far-IR. \citet{An2011} listed this object as a possible YSO based on photometric data and their analysis of the shape of its strong CO$_2$ ice feature at $\sim15.2$\,\micron. Our own data reduction does suggest an absorption component near 15.4\,\micron.  There is very strong absorption due to silicates, CO$_2$ ice at $\sim$15\,\micron, ices at 6.0 and 6.8\,\micron\ and PAH emission. There is also H$_2$ rotational line emission which is an indicator of an embedded YSO which has not yet developed an ultracompact H\,\textsc{ii} region \citep{Varricatt2010}. It appears to have the spectral characteristics of an `outflow' YSO as described by \citet{Simpson2011}.  We conclude that SSTGC~372630 is also a massive YSO.

\subsection{2MASS J06314796+0419381}

\begin{figure}
\includegraphics[angle=-90,width=\columnwidth]{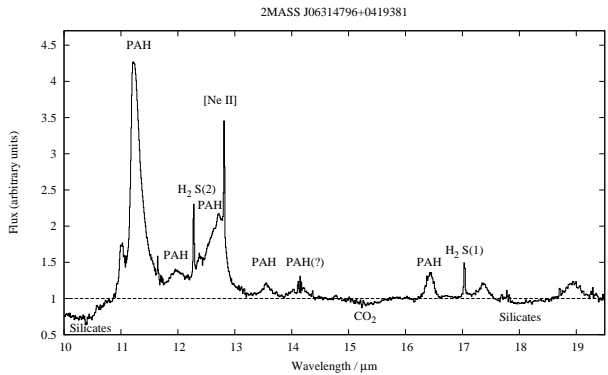}
	\caption[IRS spectrum from 5 to 19.5\,\micron\ of 2MASS J06314796+0419381] {Continuum-normalised spectrum of 2MASS J06314796+0419381 between 10 and 19.5\,\micron\ showing the presence of silicate and CO$_2$ ice absorption.}
	\label{fig:2massrosettecont}
\end{figure}

2MASS J06314796+0419381 lies in the outskirts of the Rosette nebula, which is a site of recent star formation \citep{Kuchar1993}. Invisible in optical images, this object only becomes detectable in the near-IR (see table~\ref{tab:targets}) and brightens rapidly longward of 15\,\micron. The heavy extinction of this object, together with CO$_2$ ice and silicate absorption (see figure~\ref{fig:2massrosettecont}), the presence of H$_2$ rotational emission and PAH features are supportive of it being classified as a massive YSO. There are no photometric or spectroscopic data available beyond 21\,\micron\ so the presence or absence of a 30\,\micron\ emission feature is not known.  As there are no directly relevant photometric data and the longer wavelength SED is not established, we treat 2MASS J06314796+0419381 as a \emph{candidate} massive YSO.

\subsection{YSO Evolutionary stage}
\label{sec:evolution}

The three YSO targets are at different stages of evolution, and fall between the extremes represented by Cep~A~East and S106~IR which are examples of deeply embedded and more evolved massive YSOs, respectively \citep{vandenancker2000}.  In Cep~A~East there are strong silicate and ice absorption features but no PAHs, whereas S106~IR is characterised by emission in fine-structure lines of atomic ions and from PAHs, reflecting the impact of UV radiation on the environment and its capacity to create H\,\textsc{ii} regions and PDRs. Emission from atoms and molecules excited through shocks of various types is also expected to be prevalent. In broad terms SSTGC~372630 is relatively deeply embedded, 2MASS~J06314796+0419381 is less so and ISOGAL-P J174639.6-284126 lies closer in form to the more evolved object S106~IR.  The more evolved stage of ISOGAL-P J174639.6-284126 is supported through prominent \Suiii\ (very weak in our other two objects), strong \neii\ and PAH band ratios similar to those seen in H\,\textsc{ii} regions. It is likely that both shocks and UV flux are present in the three young objects but to differing degrees.


\section{Spectra of \csixty-containing objects}
\label{sec:C60bands}

\begin{table*}
\caption[Wavelength, FWHM and flux values for the YSO targets]{Wavelength ($\lambda$, in \micron), full width at half maximum (FWHM, in \micron) and flux (in mJy\,\micron) for the 7.0, 17.4 and 18.9\,\micron\ \csixty\ emission bands in the YSO and other targets.}
 \begin{tabular}{lccccccccc}
  \hline
  Object & \multicolumn{3}{c}{7.0\,\micron} & \multicolumn{3}{c}{17.4\,\micron} & \multicolumn{3}{c}{18.9\,\micron}\\
   & $\lambda$ & FWHM & Flux\,/\,mJy\,\micron  & $\lambda$ & FWHM & Flux\,/\,mJy\,\micron & $\lambda$ & FWHM & Flux\,/\,mJy\,\micron\\
  \hline
  ISOGAL-P J174639.6-284126 & 7.01$^a$ & 0.09$^a$ & $57.7\pm1.0^a$ & 17.38 & 0.29 & $72.1\pm8.5$ & 18.91 & 0.32 & $169.5\pm7.9$\\
  SSTGC 372630 & 7.04 & 0.16 & $13.28\pm1.1$ & 17.38 & 0.34 & $12.9\pm1.3$ & 18.92 & 0.30 & $21.6\pm1.2$\\ 
  2MASS J06314796+0419381 & 7.01 & 0.11 & $6.47\pm0.14$ & 17.38 & 0.22 & $40.550\pm1.9^b$ & 18.94 & 0.30 & $61.7\pm2.7$\\ 
  HD~97300 & - & - & - & - & - & - & 18.92 & 0.27 & $61.8\pm1.7$\\
  HR~4049 & - & - & - & 17.32  & $<$1.24 & $<1239\pm28$ & 18.89 & 0.64 & $1565\pm48$\\
  HD~52961$^c$ & - & - & - & 17.39  & 0.28 & $37.5\pm2.2$ & 18.91 & 0.35 & $99.0\pm1.9$\\
  \hline
  \multicolumn{10}{l}{$^a$ Includes a contribution from an \arii\ line. $^b$ Includes a contribution from a PAH feature.  $^c$ See also \citet{Gielen2011a}.}
 \end{tabular}
\label{tab:fluxes}
\end{table*}

In this section we describe analysis of the \csixty\ emission bands in the two YSOs and candidate YSO, the Herbig~Ae/Be star HD~97300 and the post-AGB object HR~4049, spectra of which are shown in figures 4-8. For each emission feature, integrated fluxes (table~\ref{tab:fluxes}) were measured after subtraction of the nearby continuum which was fitted with a cubic spline. Whilst the \csixty\ band at 18.9\,\micron\ is an isolated feature, the other three \csixty\ bands are blended with atomic and/or PAH emission to varying degrees for each object. The 7.0\,\micron\ \csixty\ band is sometimes contaminated by an \arii\ emission line at 6.99\,\micron\, and potentially also by a band of C$_{70}$, but as this region is only available in the low resolution SL spectra deblending of these features is difficult. The 8.5\,\micron\ \csixty\ band is heavily blended with a broad PAH feature peaking at 8.6\,\micron\ which has prevented measurement of the flux in this band for all targets. The 17.4\,\micron\ \csixty\ band may include some contribution from PAH emission. None of the spectra have the strongest C$_{70}$ emission feature at 15.6\,\micron, suggesting that C$_{70}$ is either absent or present in concentrations too low to be detected, and no evidence for bands of C$_{24}$ as predicted by \citet{Kuzmin2011} was found.

The YSOs with \csixty\ bands (sections~\ref{sec:iso}-\ref{sec:2mass}) have a number of common spectral characteristics.  All have PAH features and \neii\ line emission, and two of the objects (SSTGC~372630 and 2MASS~J06314796+0419381) have pure rotational H$_2$ emission lines.  In ISOGAL-P J174639.6-284126 there are also \Siii\ (34.82\,\micron) and \Suiii\ (18.71 \& 33.48\,\micron) emission lines.

\subsection{YSO ISOGAL-P~J174639.6-284126}
\label{sec:iso}

\begin{figure}
	\includegraphics[angle=-90,width=\columnwidth]{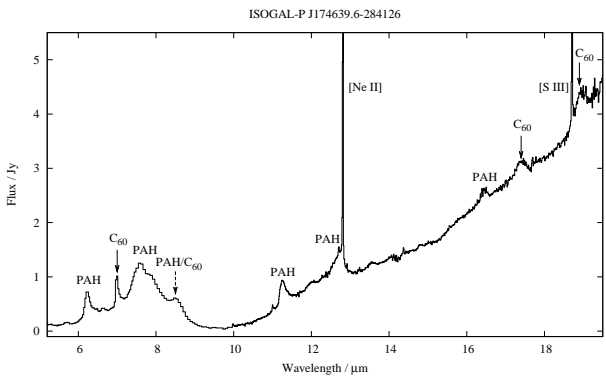}
	\caption[IRS spectrum of CMZ ISOGAL-P J174639.6-284126 in the region of the \csixty\ bands]{\emph{Spitzer} IRS spectrum of the CMZ YSO ISOGAL-P J174639.6-284126 covering the region of the \csixty\ bands at 7.0, 8.5, 17.4 and 18.9\,\micron\ \citep{Frum1991}. The expected positions of the \csixty\ bands are marked with arrows, which are solid where the feature is detected and dashed where only an upper limit on the flux is determined. PAH features at 6.2, 7.7, 8.6, 11.2, 12.7 and 16.4\,\micron\ are also marked.}
	\label{fig:isogal}
\end{figure}

The spectrum of ISOGAL-P~J174639.6-284126 covering the \csixty\ bands is shown in figure~\ref{fig:isogal} and the derived fluxes are listed in table~\ref{tab:fluxes}. In this and for all other objects the flux  quoted includes any \arii\ contribution to the 7.0\,\micron\ band. The absence of the high ionisation lines - [Ne\,\textsc{iii}] 15.55, [Ar\,\textsc{iii}] 8.99, and [P\,\textsc{iii}]  17.89\,\micron\ - which are seen in the planetary nebula Tc~1 \citep{Cami2010} suggests that the UV radiation field is less harsh than in Tc~1. However, the observed [Ne\,\textsc{ii}] 12.81 and [S\,\textsc{iii}] 18.71\,\micron\ lines do indicate the presence of a compact or ultracompact H\,\textsc{ii} region \citep{Seale2009}. The PAH spectrum, and in particular the low 11.2\,\micron/12.7\,\micron\ band ratio ($\sim$1$:$1) is also consistent with the presence of a H\,\textsc{ii} region \citep{Hony2001}.

\subsection{YSO SSTGC~372630}

\begin{figure}
	\includegraphics[angle=-90,width=\columnwidth]{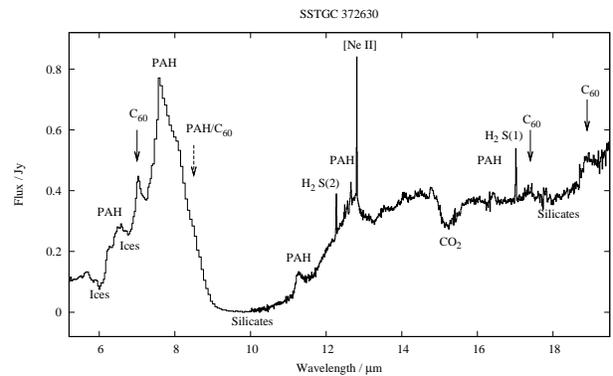}
	\caption[IRS spectrum of the CMZ YSO SSTGC~372630 in the region of the \csixty\ bands]{\emph{Spitzer} IRS spectrum of the CMZ YSO SSTGC~372630 covering the region of the \csixty\ bands. In addition to the lines identified in figure~\ref{fig:isogal}, very strong absorption due to silicates, CO$_2$ ice at $\sim$15\,\micron\ and ices at 6.0 and 6.8\,\micron\ is present. Rotational lines of H$_2$ are also marked. Note the very strong PAH emission band at 7.7\,\micron.}
	\label{fig:sstgc3}
\end{figure}

In addition to \csixty\ emission in the 18.9 and 17.4\,\micron\ bands, the spectrum of SSTGC~372630 contains \csixty\ emission at 7.0$\,\umu$m, strong silicate, CO$_2$ and other ice absorption features, and H$_2$~S(1) and S(2) rotational emission lines (see figure~\ref{fig:sstgc3}). The 7.7\,\micron\ PAH emission band is distorted by the broad silicate absorption, but still appears very strong when compared to the other PAH bands. Although the [Ne\,\textsc{ii}] line is much weaker than in ISOGAL-P~J174639.6-284126, there may still be a small contribution to the 7.0$\,\umu$m emission feature from \arii.

\subsection{YSO Candidate 2MASS~J06314796+0419381}
\label{sec:2mass}

\begin{figure}
	\includegraphics[angle=-90,width=\columnwidth]{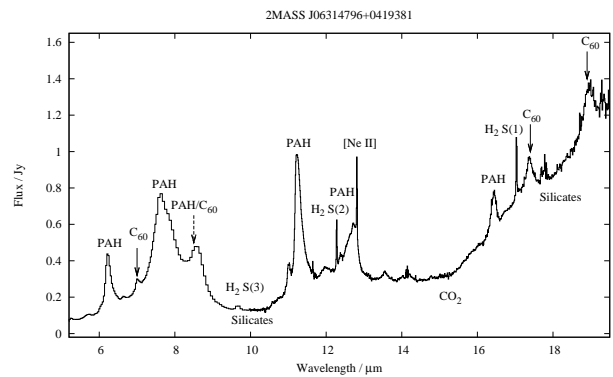}
	\caption[IRS spectrum of the Rosette Nebula YSO 2MASS~J06314796+0419381 in the region of the \csixty\ bands]{\emph{Spitzer} IRS spectrum of 2MASS~J06314796+0419381 in the Rosette nebula covering the region of the \csixty\ bands. Annotations are as in earlier figures. There is a rapid rise in flux at longer wavelengths.}
	\label{fig:2massrosette}
\end{figure}

The spectrum of 2MASS~J06314796+0419381 given in figure~\ref{fig:2massrosette} has the weakest ionised atomic lines of the three objects, with barely discernible [S\,\textsc{iii}], weak [Ne\,\textsc{ii}] and hence likely negligible \arii\ at 6.99\,\micron. Strong \csixty\ bands at 18.9, 17.4 and 7.0\,\micron\ are readily identifiable, together with PAH emission bands and the S(1), S(2) and S(3) lines of H$_2$, where the S(3) line falls in the region of very strong silicate absorption.

\subsection{Herbig Ae/Be star HD~97300}

\begin{figure}
	\includegraphics[angle=-90,width=\columnwidth]{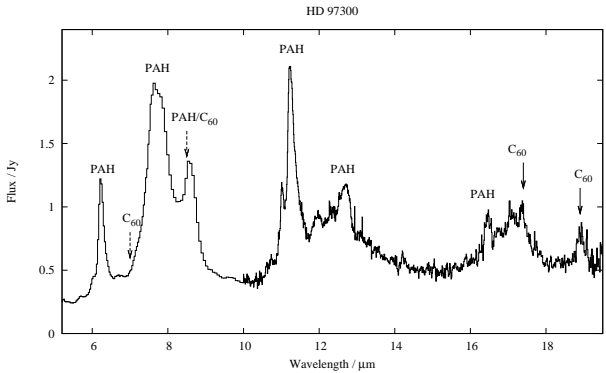}
	\caption[IRS spectrum of HD~97300 in the region of the \csixty\ bands]{\emph{Spitzer} IRS spectrum of the Herbig Ae/Be star HD~97300 covering the region of the \csixty\ bands. Annotations are as in earler figures.}
	\label{fig:hd97300}
\end{figure}

HD~97300 is an optically bright Herbig~Ae/Be star of spectral type B9. It is a more evolved system than the YSOs and is approaching the stellar main sequence \citep{Siebenmorgen1998}. The IRS spectrum is shown in figure~\ref{fig:hd97300}, where \csixty\ emission in the 18.9\,\micron\ band is visible. Emission is also seen in the region of the 17.4\,\micron\ \csixty\ band but this is heavily contaminated by PAH bands, so it is not possible to measure the flux from \csixty. No \csixty\ feature is seen at 7.0\,\micron\ suggesting a low level of vibrational excitation, corresponding to a vibrational temperature of $\leq$~200\,K \citep{Cami2011} - see also section~\ref{sec:excitation}. Weak emission in the S(1), S(2) and S(3) lines of H$_2$ is present.  Although the identification of \csixty\ in HD~97300 is based mostly on the 18.9\,\micron\ emission feature, it is reproducible as seen in other spectra \citep{Keller2008,Manoj2011}.

\subsection{Post-AGB star HR~4049}

\begin{figure}
	\includegraphics[angle=-90,width=\columnwidth]{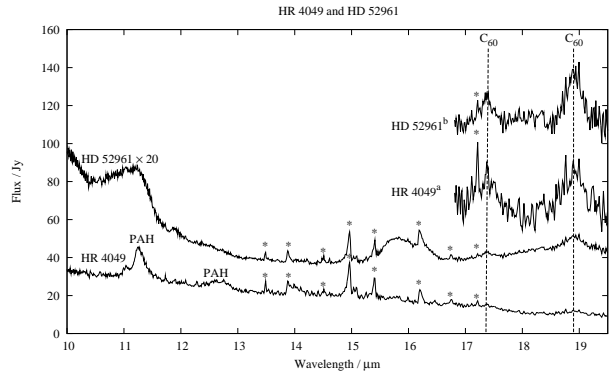}
	\caption[IRS spectra of HD~97300 in the region of the \csixty\ bands]{\emph{Spitzer} IRS spectra of the post-AGB objects HR~4049 and HD~52961. The * symbol denotes gas-phase CO$_2$ emission.  Inset: \emph{a}~$\times$~10, offset~60 and \emph{b}~$\times$~100, offset~110.}
	\label{fig:postAGBcombi2}
\end{figure}

The \emph{Spitzer} spectrum for this post-AGB object is shown in figure~\ref{fig:postAGBcombi2} together with our reduced spectrum of HD~52961, on which a study \citep{Gielen2011} has been published since this paper was submitted.  Although \citet{Gielen2011} conclude that HR~4049 does not carry the signatures of \csixty\, our analysis indicates that there is evidence of \csixty\ emission as shown in the top-right inset of the figure. Due to some contamination of the 17.4\,\micron\ \csixty\ band from gas phase CO$_2$ and PAH emission, only an upper limit on the flux in this band could be determined.

\begin{table*}
\centering
\caption{Observed vibrational band intensity ratios, inferred vibrational
temperatures for YSOs and comparison with predicted photo-excited band
ratios of \citet{Sellgren2010}.  $T_{\rm{vib}}^{\rm{C}}/K$ and
$T_{\rm{vib}}^{\rm{I-G}}/K$ are the vibrational temperatures derived from the \csixty\ band strengths used by
\citet{Cami2010} and obtained by \citet{Iglesias2011}, respectively.}
\label{tab:fluxratiosandtemeratures}

\begin{tabular}{lcccc}
\hline
Object$\dag$ & $I_{7.04}/I_{18.9}$ & $I_{17.4}/I_{18.9}$ & $T_{\rm{vib}}^{\rm{C}}/K$ & $T_{\rm{vib}}^{\rm{I-G}}/K$\\
\hline
ISOGAL-P~J174639.6-284126 (CMZ) & $\sim$\,0.42$^a$ & 0.53 & $\leq$\,450$^a$ & $\leq$\,500$^a$\\
SSTGC~372630 (CMZ) & $\leq$\,0.70$^b$ & $\sim$0.59$^b$ & $\leq$\,540$^b$ & $\leq$\,620$^b$\\
2MASS~J06314796+0419381 & 0.29 & 0.48$^c$ & 410 & 450 \\
\hline
Photon energy/eV  &  &  &   \\
\hline
 5  & 0.46-0.58 & 0.28-0.38 &   \\
10  & 0.76-0.94 & 0.28-0.38 &     \\
15  & 0.97-1.20 & 0.29-0.38 &    \\
\hline
\multicolumn{5}{l}{$^a$ Value when 60\% contribution to 7.0\,\micron\ feature from \arii\ is removed (see text).}\\
\multicolumn{5}{l}{$^b$ Silicate and ice absorptions affect continuum level definition.}\\
\multicolumn{5}{l}{$^c$ Ratio when contribution of 20\% from PAH feature at 17.4\,\micron\ is removed.}\\
\end{tabular}
\end{table*}

\section{Discussion}
\label{sec:discuss}

In this section we discuss the appearance of \csixty\ in YSOs and other objects with particular emphasis on its excitation~(section~\ref{sec:excitation}) and formation (section~\ref{sec:formation}).

\subsection{Excitation of \csixty\ IR emission bands}
\label{sec:excitation}

Two excitation mechanisms for \csixty\ IR emission with a quantitative basis have been discussed in the literature.  \citet{Cami2010} put forward a model in which thermal equilibrium is assumed, which then allows a vibrational temperature to be deduced.  In Tc~1 the temperature was found to be unexpectedly low at 330~K and led to the suggestion that the emitting \csixty\ molecules are attached to dust grains.  In their study of \csixty\ emission from NGC~2023 and NGC~7023, \citet{Sellgren2011} invoked a gas-phase photoexcitation model commonly applied in interpreting astronomical infrared emission spectra of PAH molecules. In this case absorption of a UV photon is followed by internal re-distribution of the energy leading to emission in the mid-IR.

Both of these quantitative approaches rely on knowledge of intrinsic infrared band intensities which are quite uncertain.  For the 18.9, 17.4, 8.5 and 7.0\,\micron\ bands \citet{Cami2010} used relative band strengths of 100~:~48~:~45~:~37.8 \citep{Martin1993,Fabian1996}, whereas \citet{Sellgren2010} used those of \citet{Choi2000}: 100~:~26~:~31~:~46.  Recently \citet{Iglesias2011} measured relative band intensities of 100~:~43~:~26.2~:~26.9 at room temperature in a KBr matrix. The relative band intensities are also likely to be temperature-dependent; notably emission spectra at thermal equilibrium in the gas phase at $\sim$\,1000~K reveal the 18.9 and 17.4\,\micron\ emission features to have approximately the same intensity \citep{Frum1991}.

\begin{figure}
	\includegraphics[angle=-90,width=\columnwidth]{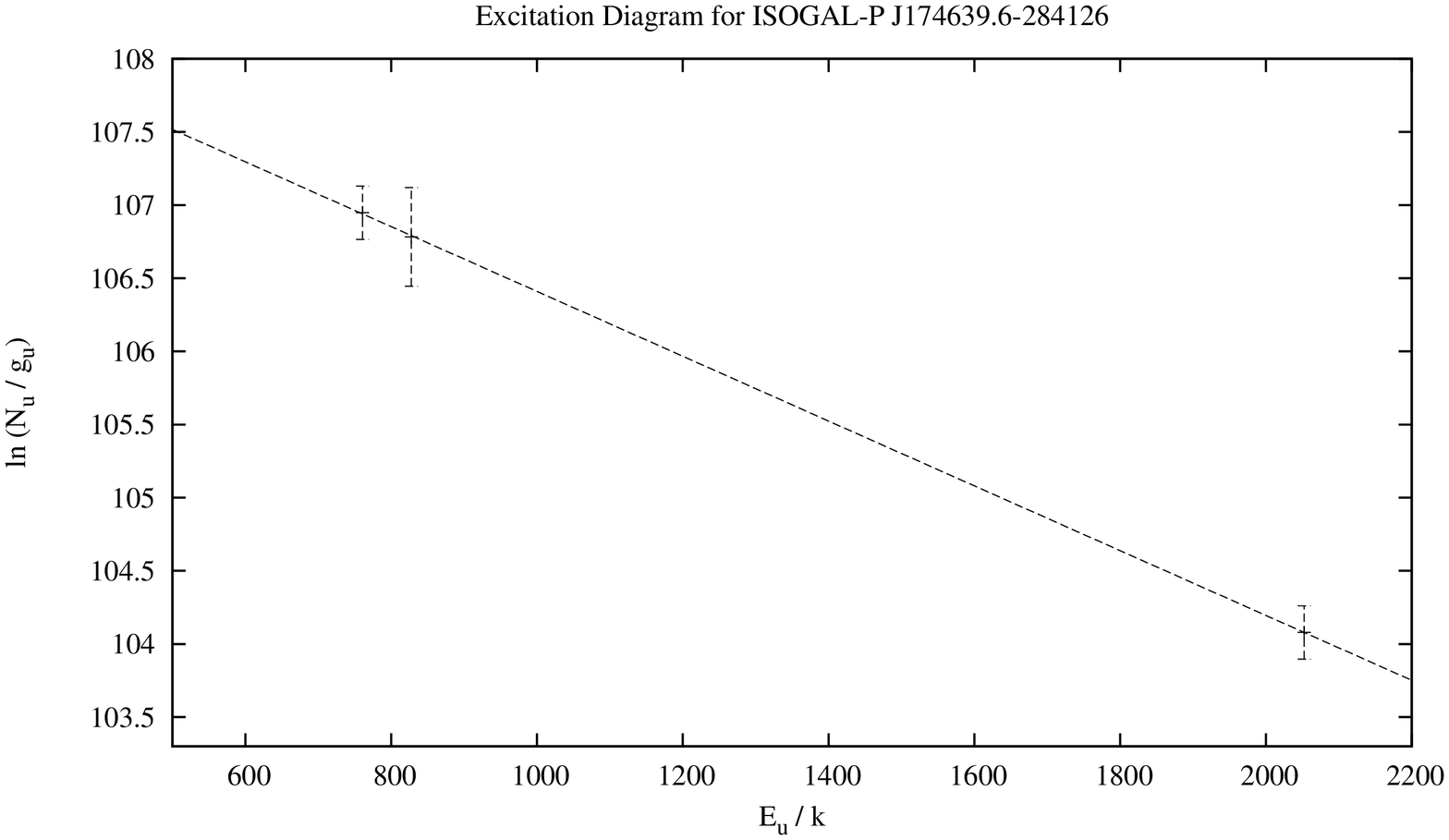}
	\caption[Excitation diagram]{Plot of ln($N_{\rm{u}}$/g$_{\rm{u}}$) \emph{versus} $E_{\rm{u}}$/k where N$_{\rm{u}}$ is the number of molecules in the upper (u) vibrational level, g$_{\rm{u}}$ is the upper level degeneracy, E$_{\rm{u}}$ is the level energy, and k is the Boltzmann constant. For details see \cite{Cami2010}. Data for the 18.9, 17.4 and 7.0\,\micron\ bands of \csixty\ in the YSO ISOGAL-P~J174639.6-284126 are shown. For this example the band strengths \citep{Martin1993,Fabian1996} were used.  A vibrational temperature of $\sim$450~K is obtained from the slope.}
	\label{fig:Tempdiagramrev}
\end{figure}

Vibrational temperatures for our objects have been estimated using the method described by \cite{Cami2010} by plotting ln($N_{\rm{u}}$/g$_{\rm{u}}$) \emph{versus} $E_{\rm{u}}$/k in an excitation diagram; an example for ISOGAL-P~J174639.6-284126 is shown in figure~\ref{fig:Tempdiagramrev}.  Using the flux values given in table~\ref{tab:fluxes}, which as listed include the \arii\ contribution to the 7.0\,\micron\ band, the \csixty\ vibrational temperature for ISOGAL-P~J174639.6-284126 is found to be $\leq670$\,K\ (using band strengths taken from \cite{Cami2010}) or $\leq790$\,K\ (using band strengths of \cite{Iglesias2011}). By applying two-component deconvolution fitting in \textsc{iraf}, the contribution of \arii\ is estimated to be $\sim60\%$ and removal leads to lower vibrational temperatures of $\leq450$\,K or $\leq500$\,K, respectively.  Vibrational temperatures for ISOGAL-P~J174639.6-284126 and the other two YSOs are given in table~\ref{tab:fluxratiosandtemeratures}. The values are in the same range as determined for PN~Tc~1 by \citet{Cami2010}, PNe by \citet{Garcia2010} and a PPN by \citet{Zhang2011}, although the low signal-to-noise ratio of the \csixty\ bands in SSTGC~372630 renders the apparently slightly higher temperature values in this YSO less certain.   We have not conducted our own photexcitation calculations for infrared emission from \csixty, but predicted infrared band intensity ratios for photoexcitation energies of 5, 10 and 15~eV are available from \citet{Sellgren2010} and are given in table~\ref{tab:fluxratiosandtemeratures}.

We now consider whether either of the thermal or photoexcitation models hold for these YSOs.  Examination of table~\ref{tab:fluxratiosandtemeratures} shows that, with one possible exception ($I_{7.04}/I_{18.9}$ for SSTGC~372630), the $I_{7.04}/I_{18.9}$ and $I_{17.4}/I_{18.9}$ ratios fall outside the ranges calculated by \citet{Sellgren2010} for photons with 5, 10 and 15~eV.  If the \csixty\ molecule were excited by UV radiation, the flux ratios derived for ISOGAL-P J174639.6-284126 and 2MASS~J06314796+0419381 (see table~\ref{tab:fluxratiosandtemeratures}) would suggest the need for photons with energies lower than 5~eV.  A similar result is found for Tc~1 \citep{Cami2011} and for a number of PNe \citep{Garcia2011a}.  However, for the PPN~IRAS~01005+7910 the band ratios can be satisfied within this model but with high-energy photons of around 10~-~15~eV \citep{Zhang2011}. The case of the pre-main sequence Herbig Ae/Be star HD~97300 is notable because even for photon energies as low as 5 eV, a ratio of 0.46-0.58 for the 7.0/18.9 band ratio is predicted. While the 18.9\,\micron\ band of \csixty\ is clearly discernible in HD~97300, a band at 7.0\,\micron\ is not seen which implies emission from very cool \csixty. Unfortunately the 17.4\,\micron\ feature is too heavily contaminated by PAH features to allow a value for $T_{\rm{vib}}$ to be deduced.

From these comparisons, an emission mechanism based solely on the absorption of UV radiation followed by re-emission in the infrared would seem unlikely for these four new \csixty-containing young objects.   However, within the thermal model, the derived temperatures (table~\ref{tab:fluxratiosandtemeratures}) for the three YSO objects are not inconsistent with the temperatures of up to $\sim$1000~K for warm gas in massive YSOs \citep{Lahuis2000,Boonman2003,An2011}. While the location of the \csixty\ in the YSOs is not yet established, and there is no evidence for gas-phase absorption by HCN, C$_2$H$_2$ or CO$_2$ in the spectra, warm gaseous regions could provide a suitable environment for thermal excitation of IR emission from \csixty.

For SSTGC~372630 (CMZ) and 2MASS J06314796+0419381 (Rosette nebula) there are S(1) and S(2) pure rotational lines of molecular hydrogen.  Assuming that the lines are optically thin and that thermal equilibrium holds both for the rotational levels and for the ortho/para ratio, then rotational temperatures can be deduced.  We find \emph{T}$_{\rm{rot}}$ values of $\sim$540 and $\sim$370~K for SSTGC~372630 and 2MASS~J06314796+0419381 which are within $\sim$~100~K of the vibrational temperatures for \csixty\ in these objects. In addition to the assumptions listed, this difference may have a number of contributing factors including the molecular physics of the excitation, the allowed vibrational transitions of \csixty\ and the strongly forbidden rotational ones of H$_2$, and uncertainty as to whether \csixty\ and H$_2$ share the same spatial distribution.  Nevertheless, to our knowledge this is the first time that a \csixty\ vibrational temperature has been compared with the internal temperature of another molecule. Observations of other (polar) molecules through their rotational spectra in these objects would clearly be of interest.

Further possible \csixty\ vibrational excitation mechanisms include shock-induced excitation, possibly during \csixty\ formation from HAC in grain-grain collisions, or through dehydrogenation of HAC while on the grain surface \citep{Garcia2011a,Cami2011}.  The YSOs SSTGC~372630 and 2MASS~J06314796+0419381, show H$_2$ S(1) and S(2) lines which is an indicator of shocked gas, while all three YSOs have strong \neii, also an indicator of shocked gas, but no \neiii\ emission. In their study of \csixty-containing PNe, \citet{Garcia2011a} found very low \neiii/\neii ratios. This result was considered incompatible with the predictions of photoionisation modelling, but one that could possibly be rationalised in terms of shocks.  Shocks are expected in YSOs and so could provide \csixty\ vibrational excitation.  In the case of SSTGC~372630, the H$_2$ lines are shifted by $\sim$200~kms$^{-1}$ relative to the \neii\ {line, strongly suggesting a strong outflow/shock which is consistent with the likelihood that this object is an `outflow' source as mentioned in section~\ref{sec:sstspectrum}.\\

\subsection{Formation of \csixty}
\label{sec:formation}

It is well established that \csixty\ is formed in the later stages of stellar evolution (see \cite{Cami2010}, \cite{Garcia2010} and references in section~\ref{sec:intro}). Its presence in NGC~2023 and NGC~7023 \citep{Sellgren2010} indicates that it might also be formed \emph{in situ} in the ISM, where it may also exist in ionised form \citep{Foing1994}. Discovery of \csixty\ in the YSOs reported here makes the question as to the mechanism and location for \csixty\ formation an even wider one. In this section we discuss how \csixty\ could form \emph{in situ} in pre-main-sequence objects or arise from earlier synthesis in post-AGB stars or the ISM (section~\ref{sec:csixtyinysos}), and draw attention to a tentative link between \csixty\ and nanodiamonds (section~\ref{sec:nanodiamonds}).


\subsubsection{\csixty\ in the YSOs and Herbig Ae/Be star HD~97300}
\label{sec:csixtyinysos}

A number of scenarios for the origin of \csixty\ in YSOs can be envisaged. Given that it is clearly formed in post-AGB objects, then if the integrity of the \csixty\ structure were maintained on it being expelled to the ISM, it could then (re)appear in the spectrum of a YSO following cloud collapse.  For a discussion on the evolution of organic material through these stages, see \citet{Ehrenfreund2000}.  Alternatively \csixty\ in YSOs could originally have formed in the ISM as discussed by \citet{Bettens1996,Bettens1997}. A third scenario is that \csixty\ forms \emph{in situ} in YSOs.  Recent commentaries on the current challenges relating to both mechanistic and spatial aspects of \csixty\ formation have been published by \citet{Cami2011} and by \citet{Garcia2011b}.  Attention is focussed here on \emph{in situ} formation of \csixty\ in YSOs.

We consider two main schemes. Firstly, shock-induced formation of \csixty\ might occur \emph{in situ} in YSOs through decomposition of HAC, essentially as described by \citet{Garcia2010} for post-AGB objects. However, it is not clear whether HAC is present in YSOs. For post-AGB objects there is a very strong correlation between \csixty\ and the broad 30\,\micron\ feature that is commonly attributed to HAC.  However, the \csixty-containing reflection nebulae NGC~2023 and NGC~7023 do not have the 30\,\micron\ emission (see \cite{Zhang2011}), and nor do the YSOs SSTGC~372630 and ISOGAL-P J174639.6-284126, the Herbig Ae/Be star HD~97300 \citep{Keller2008} or the mixed-chemistry post-AGB objects HR~4049 and HD~52961. Hence, if the 30\,\micron\ emission does indeed arise from HAC then it is hard to see how this particular shock-induced \csixty-formation mechanism involving HAC could be operating in young stellar environments.\footnote{The anonymous referee has pointed out that the absence of a 30\,\micron\ feature could be related to the temperature of the bulk of the HAC material.}

Secondly, in most astrophysical sources where \csixty\ has been found, PAH emission is also present. This raises the question as to whether there might be a link between PAHs (perhaps particularly in dehydrogenated form) and fullerene formation. \cite{Micelotta2010a,Micelotta2010b} have argued from a theoretical standpoint that fullerene formation could occur through shock driven processing of PAHs.  They conclude that shock velocities of 75-100~kms$^{-1}$ can significantly modify the structure of PAHs, these speeds being comparable to those found in massive YSOs. Even at velocities as low as 50~kms$^{-1}$ some carbon atoms are expected to be removed from the carbon framework as could occur in C-type shocks (e.g. \cite{vandenancker2000}).  \citet{Chuvilin2010} have recently shown that under highly energetic conditions small PAH-like graphene sheets can undergo re-arrangement to form fullerenes.\footnote{Since this paper was submitted an arXiv manuscript by \cite{Berne2011} has appeared which discusses this work with reference to NGC~7023.}  Although the laboratory experiments of \cite{Chuvilin2010} were conducted using transmission electron microscopy in which very high energy electrons induce dehydrogenation and loss of carbon atoms leading to \csixty\ formation, high energy photons, x-rays and cosmic rays could play an equivalent role in astrophysical environments. These processes all proceed via initial dehydrogenation and loss of carbon atoms leading to formation of 5-membered rings. Curvature follows with the final stage being closure of the fullerene structure. Given that PAHs are present in all of the YSOs, `top-down' mechanisms starting from large PAHs would appear to be favoured as formation routes for \csixty\ in YSOs as well as in the Herbig Ae/Be star HD~97300.

\subsubsection{\csixty\ in post-AGB stars: HR~4049 and HD~52961}
\label{sec:nanodiamonds}

During the search of \emph{Spitzer} archival data two post-AGB sources with \csixty\ emission at 18.9\,\micron\ were found, HR~4049 and HD~52961~ (see table~\ref{tab:targets} and figure~\ref{fig:postAGBcombi2}). These objects are unusual in that they have a mixed oxygen-carbon chemistry and so the presence of \csixty\ is perhaps unexpected. The presence of a band of \csixty\ in HD~52961 has also been recently reported by \citet{Gielen2011}.\footnote{Since this paper was submitted an arXiv manuscript by \cite{Gielen2011a} has appeared in which a discussion of possible formation mechanisms in a mixed oxygen-carbon environment is presented.} We include these objects because of the parallel between the outflows (jets and shocks) that are common in the later stages of stellar evolution and those occurring in young objects. It follows that \csixty\ formation processes in these two very different evolutionary stages may be quite similar.  We suggest that `top-down' formation of \csixty\ from PAHs as described by \cite{Micelotta2010a,Micelotta2010b} and seen experimentally by \cite{Chuvilin2010} will likely apply also to \csixty\ formation in mixed chemistry post-AGB objects such as HR~4049.

HR~4049 and HR~52961 exhibit the very rare C--H emission features at 3.43/3.53\,\micron\ from H-coated nanodiamonds \citep{Geballe1989,Oudmaijer1995,Guillois1999}. Given that they also have \csixty\ emission, this suggests that there may be a connection between fullerenes and nanodiamonds. Carbon onions are composed of fullerene-like concentric shells and due to their similarity with fullerenes may form together in astrophysical environments. It has been proposed that astrophysical carbon onions could act as pressure cells for the formation of nanodiamonds under highly energetic conditions \citep{Goto2009}. The fact that the only post-AGB objects with nanodiamond signatures also have \csixty\ bands provides indirect observational support for the nanodiamond formation proposal of \cite{Goto2009}.

\section{Summary}
\label{sec:summary}

This paper reports the first detection of the \csixty\ molecule in three young stellar objects and in a Herbig Ae/Be star using \emph{Spitzer} IRS archive observations. Evidence that the three objects are YSOs is critically reviewed and their relative stage of evolution discussed.   \csixty\ emission in two unusual mixed-chemistry post-AGB objects is also described as they share shock-induced processes similar to those active in YSOs.  A common feature of all of the objects is emission from PAHs.

The origin of the \csixty\ emission is considered in detail.  From the measured fluxes in the 7.0, 17.4 and 18.9\,\micron\ bands, it is found that the band intensities can best be described by a thermal rather than a UV photoexcitation model and approximate vibrational temperatures were determined.  These are similar to the rotational temperatures deduced from H$_2$ line emission found in two of the young objects.

 \csixty\ formation mechanisms are evaluated, focussing particularly on photo-induced and shock-induced \emph{in situ} formation from PAHs in YSOs.  We suggest the most likely formation mechanism is through dehydrogenation followed by carbon-atom loss from PAHs resulting in five-membered ring formation, curvature and closure to make fullerenes.  It is concluded that shock-induced and UV stimulated decomposition are active in generating \csixty\ in the YSOs, with the relative contributions varying with evolutionary stage. It is thought that shocks play the dominant role in \csixty\ formation in the Herbig Ae/Be and post-AGB stars.

\section*{Acknowledgments}

KRGR thanks EPSRC for a studentship. This work is based on observations made with the \emph{Spitzer} Space Telescope, which is operated by the Jet Propulsion Laboratory, California Institute of Technology under a contract with NASA.

\bibliographystyle{mn2e}
\bibliography{c60}

\bsp

\label{lastpage}

\end{document}